\documentclass[11pt, a4paper, logo, onecolumn, copyright]{googledeepmind}

\usepackage[authoryear, sort&compress, round]{natbib}

\usepackage{times}
\usepackage{soul}
\usepackage{url}

\usepackage[ruled,linesnumbered]{algorithm2e}
\usepackage{algpseudocode}
\usepackage{amsfonts}
\usepackage{balance}
\usepackage[table]{xcolor}
\usepackage[separate-uncertainty,table-align-uncertainty,retain-zero-uncertainty]{siunitx}
\usepackage{tablefootnote}
\usepackage{listings}  
\usepackage{enumitem} 
\definecolor{codegray}{gray}{0.95}
\definecolor{commentgreen}{rgb}{0.13, 0.54, 0.13}
\definecolor{keywordblue}{rgb}{0.13, 0.13, 1}

{}
{}

\setlist[enumerate]{nosep, wide, labelwidth=0pt, labelindent=0pt}

\setlist[itemize]{nosep, wide, labelwidth=0pt, leftmargin=0pt, labelindent=0pt}

\newlength\kwinlen

\usepackage[norule,symbol,perpage]{footmisc}

\title{Code-Space Response Oracles: Generating Interpretable Multi-Agent Policies with Large Language Models}

\correspondingauthor{hennes@google.com, lizun@google.com}

\reportnumber{} 

\author[*1]{Daniel Hennes}
\author[*1]{Zun Li}
\author[1]{John Schultz}
\author[1]{Marc Lanctot}

\affil[*]{Equal contributions}
\affil[1]{Google DeepMind}

\begin{abstract}
Recent advances in multi-agent reinforcement learning, particularly Policy-Space Response Oracles (PSRO), have enabled the computation of approximate game-theoretic equilibria in increasingly complex domains. 
However, these methods rely on deep reinforcement learning oracles that produce `black-box' neural network policies, making them difficult to interpret, trust or debug.
We introduce Code-Space Response Oracles (CSRO), a novel framework that addresses this challenge by replacing RL oracles with Large Language Models (LLMs). 
CSRO reframes the best response computation as a code generation task, prompting an LLM to generate policies directly as human-readable code. 
This approach not only yields inherently interpretable policies but also leverages the LLM's pretrained knowledge to discover complex, human-like strategies.
We explore multiple ways to construct and enhance an LLM-based oracle: zero-shot prompting, iterative refinement and \emph{AlphaEvolve}, a distributed LLM-based evolutionary system.
We demonstrate that CSRO achieves performance competitive with baselines while producing a diverse set of explainable policies. 
Our work presents a new perspective on multi-agent learning, shifting the focus from optimizing opaque policy parameters to synthesizing interpretable algorithmic behavior.
\end{abstract}

\begin{document}

\maketitle

\section{Introduction}
\label{sec:introduction}

The strategic interaction of multiple autonomous agents is a cornerstone of artificial intelligence, with applications ranging from economic modeling and autonomous driving to cybersecurity and recreational games~\citep{ShohamLeytonBrown2008}. A central goal in this field is to develop algorithms that compute robust and effective strategies. The predominant theoretical framework for analyzing such systems is game theory. However, computing exact equilibria in large, complex games is often intractable, motivating the development of iterative methods to find approximate solutions.


Among the most successful of these methods are Policy-Space Response Oracles (PSRO)~\citep{Lanctot2017PSRO,bighashdel2024policy}. PSRO and its variants have achieved state-of-the-art performance in notoriously difficult games, such as Barrage Stratego~\citep{mcaleer2020pipelinepsro} and  StarCraft~\citep{Vinyals2019AlphaStar}. PSRO iteratively builds a set of policies by computing a best response to the current meta-strategy. Standard implementations rely on deep reinforcement learning (RL) oracles, which train a neural network to approximate a best response through extensive interaction with a population of opponents. While powerful, this approach results in policies that are opaque, ``black-box'' models. This lack of interpretability prevents strategy verification, and forms a significant barrier to deploying such agents in high-stakes, real-world applications where explainability is crucial. Furthermore, training these RL oracles is often sample-inefficient, requiring millions or billions of game simulations to converge.


To address the trade-off between performance and interpretability, we propose \textbf{Code-Space Response Oracles (CSRO)}. We reframe best-response computation as a program synthesis task, replacing the Deep RL oracle with a Large Language Model (LLM) that generates policies as executable source code. By providing the LLM with the game rules, an API for interacting with the environment, and descriptions of opponent strategies, we prompt it to write a programmatic best response. 
CSRO leverages the vast knowledge of logic, planning, and strategy embedded within pre-trained LLMs, enabling them to generate sophisticated, human-like policies, dramatically reducing the need for costly exploration.

CSRO generalizes a recent demonstration called LLM-PSRO~\citep{bachrach2025combining}. LLM-PSRO relies on open-loop ``best-of-N'' sampling using full opponent source code, CSRO introduces two critical extensions for scalability and robustness. First, we replace passive sampling with an \emph{iterative refinement loop} (e.g., via evolutionary algorithms) that actively optimizes strategies based on meta-game performance. Second, we address context limits via \emph{context abstraction}; rather than ingesting raw code from all opponents, CSRO summarizes strategies in natural language and filters inputs. This allows our method to scale to complex games where full source code would exceed context windows.


This paper makes several contributions, primarily introducing CSRO as a novel framework utilizing LLMs as code-generating oracles to compute approximate equilibria. We demonstrate that while LLMs can generate strategies zero-shot, performance is significantly enhanced by integrating evolutionary refinement method, such as AlphaEvolve~\citep{romera2024mathematical,novikov2025alphaevolve}, to iteratively optimize code based on closed-loop feedback. This approach yields policies that are fully interpretable, represented by commented source code. Finally, unlike prior work limited to internal comparisons~\citep{bachrach2025combining}, we rigorously validate CSRO by benchmarking against standardized external populations and game-theoretic solvers, showing that code-generation oracles can compete with mature baselines in established domains.


\section{Code Space Response Oracles}
\label{sec:methodology}

In this section, we first provide a brief overview of the standard PSRO algorithm, which forms the foundation of our work. We then formally introduce our proposed method, CSRO, detailing how an LLM is employed as a code-generating oracle to produce interpretable, programmatic policies.

\subsection{Preliminaries}
\label{ssec:psro}

PSRO is an iterative algorithm for computing approximate Nash equilibria in multi-agent games.
In this paper we focus on two-player symmetric zero-sum games, but the methodology directly transfers to general game domains.
Consider a two-player symmetric game where each player $i \in \{0, 1\}$ shares a common set of strategies $\Pi$ with the same utility function $u$.
A strategy, or a policy, $\pi\in\Pi$, is a mapping from observations to a probability distribution over actions. 
A player playing $\pi$ receives $u(\pi, \pi')$ against an opponent playing $\pi'$; the opponent receives $u(\pi', \pi)=-u(\pi, \pi')$ according to the zero-sum condition.
The PSRO algorithm in a symmetric game setting maintains a single set of policies $P$. 
At each iteration $k$, the algorithm first solves for a symmetric equilibrium mixture $\sigma$ over the current policies in an empirical meta-game. 
Then, from the viewpoint of any player (since they are symmetric), a best response oracle computes a policy $\pi^*$ that maximizes expected utility against the opponent meta-strategy~$\sigma$:
\begin{equation}
    \label{eq:best_response}
    \pi^* \in \operatorname*{arg\,max}_{\pi} \mathbb{E}_{\pi' \sim \sigma} [u(\pi, \pi')]
\end{equation}
This new policy is added to the policy set $P$, and the process repeats. In standard PSRO, the oracle is a deep reinforcement learning algorithm that produces an opaque neural network policy.

\subsection{Code Policies}
\label{ssec:csro}

The central innovation of CSRO is the replacement of the deep RL oracle with an LLM that synthesizes policies as executable code. This reframes the best response computation from a process of numerical optimization to one of programmatic reasoning and generation. In our framework, a policy $\pi$ is no longer an opaque neural network but a (stateful) code policy, such as a Python function, that maps observations to actions.

The primary mechanism for guiding the LLM is a carefully constructed prompt, which is dynamically generated at each iteration of the algorithm. This prompt provides the LLM with all necessary context to formulate a strategy. It begins with a natural language description of the game's rules and objectives, alongside a precise API specification for the policy function to ensure the generated code is executable. A crucial component is the description of the opponent's current meta-strategy, $\sigma$. Since opponent policies in CSRO are also programmatic, their source code can be directly included in the prompt. To manage prompt complexity when facing a large mixture of policies, we can in addition to --- or instead of the code --- provide a high-level summary of the opponents' collective behavior, a summary that can itself be generated by prompting the LLM to analyze the set of code policies. The prompt concludes with a clear directive to generate a best response and can optionally include an instruction to produce self-explaining code with detailed comments and a docstring describing the intended strategy.

\begin{figure*}[!t]
    \centering
    \includegraphics[width=\linewidth,trim={.2cm .2cm .2cm .2cm},clip]{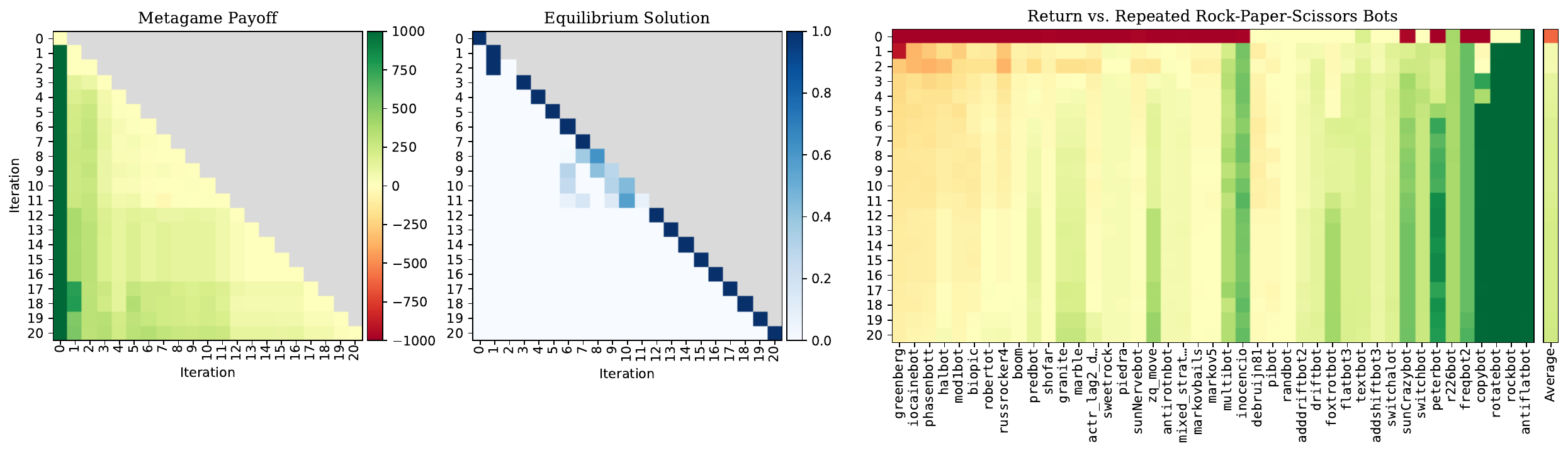}\vspace{-.5em}
    \caption{Example run of CSRO in Repeated Rock-Paper-Scissors.}\label{fig:best_return_rrps}\vspace{-.5em}
\end{figure*}

\subsection{CSRO Algorithm}
\label{ssec:csro_algorithm}

The complete CSRO algorithm is detailed in Algorithm~\ref{alg:symmetric_csro_refined}. The overall structure follows the iterative loop of PSRO, but the oracle step is replaced by an LLM-based code generation process. At each iteration, a new programmatic policy is generated, and added to the policy set, enriching the meta-game with increasingly sophisticated strategies.

\begin{algorithm}[h]



    


\DontPrintSemicolon
\LinesNumbered
\vspace{.25em}
\KwIn{$G$: symmetric zero-sum game, $K$:~max.~iterations,
$M$:~max.~refinement budget}

\BlankLine
$P \gets \{\pi_{\text{initial}}\}\;$

\For{$k \leftarrow 1$ \KwTo $K$}{
    $U \gets \texttt{compute\_payoff\_matrix}(P)$\;
    $\sigma \gets \texttt{compute\_meta\_equilibrium}(U)$\;
    
    $prompt \gets \texttt{construct\_prompt}(G, \sigma, P)$\;
    $\pi' \gets \texttt{llm\_oracle}(prompt)$\;
    
    $u \gets \texttt{evaluate\_policy}(\pi', \sigma, P)$\;
    $j \gets 0$\;

    \While{\textbf{\texttt{not}} \texttt{terminated}(u, j, M)}{
        $j \gets j + 1$\;
        $prompt \gets \texttt{update\_prompt}(prompt, \pi', u)$\;
        $\pi' \gets \texttt{llm\_oracle}(prompt)$\;
        $u \gets \texttt{evaluate\_policy}(\pi', \sigma, P)$\;
    }
    
    $P \gets P \cup \{\pi'\}\;$
}
\KwRet{$P, \sigma$}\;
\caption{\textbf{Code-Space Response Oracles.}}\label{alg:symmetric_csro_refined}
\end{algorithm}

\subsection{Oracle Refinement Mechanisms}
\label{ssec:refinement}

The CSRO framework supports two distinct, complementary mechanisms for improving the quality of generated policies using in-context learning, as detailed in Algorithm~\ref{alg:symmetric_csro_refined}. These mechanisms operate on different timescales to provide both immediate tactical corrections and long-term strategic guidance.

\subsubsection{Cross-Iteration Strategic Adaptation}
At the beginning of each iteration $k$, the oracle is provided with deep strategic context by analyzing the programmatic policies that constitute the current meta-game equilibrium $\sigma$. The \texttt{construct\_prompt} function inspects the source code of the opponent policies currently active in the meta-game (i.e., those with a non-zero probability in $\sigma$). The \emph{source code} of each opponent can be directly included in the prompt or alternatively summarized by another LLM call to provide a high-level \emph{description} of the strategic behavior the opponent code implements. To keep the context length manageable we can, e.g., filter by a threshold on \emph{minimum support} or only include the \emph{top-k} opponents according to the meta-game equilibrium. The provided opponents' source code or descriptions guide the generation of the next candidate policy. 

\subsubsection{Intra-Iteration Policy Refinement}\label{subsec:policy_refinement}
The second mechanism is an inner feedback loop designed to ensure that any new policy $\pi'$ is a robust best response to the current meta-game $\sigma$. 
We here consider three variants that differ in how the refinement stage in lines 9-14 are implemented:
\begin{enumerate}
    \item  \emph{ZeroShot}: We prompt the LLM to directly generate a program in zero-shot fashion.
    This corresponds to the case where the refinement stage (Algorithm~\ref{alg:symmetric_csro_refined}: line 9--14) is never entered.
    \item \emph{LinearRefinement}: We start with an initial candidate policy whose expected utility $u$ against $\sigma$ is immediately evaluated. If the policy is found to be losing ($u < 0$), a conditional refinement loop is triggered. The LLM is then tasked with regenerating the policy based on this feedback. The program is updated if the score is increased. This cycle of evaluation and refinement continues until the policy achieves a non-negative utility or a maximum refinement budget $M$ is reached.
    We call it "linear refinement" because all the computations are exercised in a single thread.
    \item \emph{AlphaEvolve}: AlphaEvolve~\citep{romera2024mathematical,novikov2025alphaevolve} is a large-scale distributed system which uses an LLM to mutate programs in multiple threads and uses a score function to guide an evolutionary search procedure.
    To ensure diversity, programs are clustered into different subpopulations which are evolved independently. 
    Each evolution thread continuously samples past programs and prompts the LLM to generate new modifications to them in favor of high-scored variants.
    AlphaEvolve has been shown effective in finding novel and better algorithms in math, optimization~\citep{novikov2025alphaevolve} and cognitive science domains~\citep{castrodiscovering}.
    We consider AlphaEvolve perfectly suitable for our purpose of computing best response programs using an estimate of $\mathbb{E}_{\pi' \sim \sigma} [u(\pi, \pi')]$ as the score function for the evolutionary search.
\end{enumerate}


\section{Experiments}
\label{sec:experiments}

To validate the effectiveness and interpretability of CSRO, we conduct a series of experiments designed to answer the following research questions:
(1) Does CSRO converge to a low-exploitability equilibrium? 
(2) Can the LLM oracle generate effective, non-trivial strategies in a zero-shot or few-shot manner?
(3) Are the programmatic policies generated by CSRO demonstrably more interpretable than the neural network policies produced by traditional oracles?


\subsection{Environments}
We evaluate our framework on two standard analytical games, chosen to test different aspects of strategic reasoning. Both environment implementations are based on OpenSpiel~\citep{lanctot2019openspiel}.

\vspace{.5em}\noindent\textbf{Repeated Rock-Paper-Scissors (RRPS):} In this repeated version, players repeat the standard stage game for \num{1000} consecutive rounds (\emph{throws}). Although playing according to a uniform random policy at each round is known to be a Nash equilibrium and is unexploitable, it fails to exploit patterns in the moves of suboptimal opponents. Therefore, RRPS has been a very informative benchmark for multiagent learning and opponent modeling~\citep{lanctot2023population,billings2000first,billings2000second,rpscontest}.

\vspace{.5em}\noindent\textbf{Repeated Leduc hold'em poker:} In this environment, players play Leduc hold'em for 100 rounds (\emph{hands}); the dealer role alternates between consecutive rounds, with the first-round-dealer randomly decided between the two players. 
Leduc hold'em is a popular benchmark for research in imperfect information games, introduced in~\citep{Southey2005LeducPoker}. It is a two-player, zero-sum game with a small six-card deck (two suits of Jack, Queen, and King). The game features two betting rounds, with a single public community card revealed at the flop. Its structure, which combines private and public information across multiple decision points, requires more sophisticated stateful reasoning than simpler variants like Kuhn poker, making it an excellent environment to test the LLM's ability to generate complex, multi-step strategies involving concepts like bluffing and value betting.
The repeated version of Leduc hold'em is a more realistic setting than the one-shot version; and it further adds another strategic layer where an agent is more effective if it identifies strategic patterns of the opponent and is thus able to exploit them (similar to RRPS).

\subsection{Evaluation Population}
In RRPS, we evaluate CSRO code-policies against \num{43} hand-coded, \emph{heuristic strategies} from the international RRPS competitions~\citep{billings2000first,billings2000second}. These bots range from simple, stateless strategies like \texttt{randbot} (uniform random) and \texttt{rockbot} (always rock) to bots playing fixed sequences like \texttt{pibot}. Other approaches use limited memory, such as \texttt{copybot} (beating the opponent's last move) or \texttt{driftbot} (biasing actions based on recent play). More sophisticated entrants rely on historical observations, using methods like frequency analysis (\texttt{freqbot2}), Markov chain predictions (\texttt{markov5}), or complex rule-based systems (\texttt{inocencio}). Some bots switch between different internal policies based on profitability (\texttt{multibot}), while innovative designs include a neural network-like system (\texttt{sunNervebot}) and a cognitive architecture (\texttt{actr\_lag2\_decay}). The competition winners, \texttt{iocainebot} and \texttt{greenberg}, succeeded by maintaining a large set of different predictors and counter-strategies, selecting the one that proved most effective as the match progressed.
This population of bots represent a diverse set of RRPS strategies which is useful in terms of measuring the generalization and exploitability of agents~\citep{lanctot2023population}. An example run of CSRO in the RRPS environment is visualized in Figure~\ref{fig:best_return_rrps}, showing the evolution of payoffs and the equilibrium solution over iterations as well as returns against the bot population.

For repeated Leduc Poker, we compute a Nash equilibrium for the single-hand game using \emph{Counterfactual Regret Minimization\,+} (CFR+)~\citep{tammelin2014solving}. We run CFR+ for \num{10000} iterations and use the weighted averaged strategy. We evaluate CSRO code-policies against an opponent which plays this optimal Nash strategy at every hand of the repeated game. We also evaluate code-policies against two heuristic strategies with easy to detect patterns: \emph{AlwaysCall} and \emph{AlwaysFold} (a strategy that always folds when possible and otherwise calls). 

Given an evaluation population of heuristic strategies (bots), we can adopt the three metrics introduced in~\citep{lanctot2023population}. 
We denote the population of bots as $\mathcal{P}$. 

\noindent\emph{Population Return} is the average return of $\pi$ against the population. This measures the generalization capability of agents against an unseen opponent at test-time:
$$\textsc{PopReturn}(\pi)=\mathbb{E}_{\pi'\sim\mathcal{P}}[u(\pi, \pi')]$$

\noindent\emph{Within Population Exploitability} serves as an approximate exploitability metric:
$$\textsc{PopExpl}(\pi)=-\min_{\pi'\in\mathcal{P}}u(\pi, \pi')$$

\noindent\emph{Aggregate Score} measures a balance between average- and worst-case analysis:
$$\textsc{AggScore}(\pi)=\textsc{PopReturn}(\pi)-\textsc{PopExpl}(\pi)$$
An agent can have both high $\textsc{PopReturn}$ and $\textsc{PopExpl}$ by winning against most of the bots in the population but being particularly exploitable by one opponent. In this case $\textsc{AggScore}$ is a better measure of generalization capability.

\subsection{Baselines}
To evaluate CSRO, we compare against a strong, conventional baseline from the multi-agent learning literature. We implement a standard PSRO algorithm where the best response oracle is a deep reinforcement learning agent. The agent's policy is represented by a deep Long Short-Term Memory (LSTM)-based network and is trained using the Importance Weighted Actor-Learner Architecture (IMPALA) algorithm~\citep{espeholt2018impala}.
The recurrent nature of LSTM makes it suitable for our repeated, partially observable game settings.
This serves as a canonical baseline, allowing for a direct comparison between our method, which generates oracle policies as explicit code, and the standard approach of optimizing a policy in a neural network's weight space. We refer to the Supplementary Materials~\ref{sec:psro-impala} for implementation details.

In RRPS, we further include a direct comparison with an \emph{LLM-based agent} adapting the simple yet effective approach proposed in~\citep{lanctot2023population}. The agent's architecture requires no task-specific training, instead relying directly on the sequence-prediction capabilities of a pre-trained LLM. A history of previous actions is provided as a textual prompt and the model generates the subsequent actions for both player and opponent. The agent plays the best response to the predicted opponent action. While the original work utilized Chinchilla~\citep{hoffmann2022chinchilla}, our implementation substitutes this family of LLMs for the pre-trained Gemma 3 models~\citep{gemmateam2025gemma3} to benchmark a more recent and accessible model on this task. See Table~\ref{tab:gemma3_rrps} for full results.

In addition to the LLM agent based on Chinchilla, we include two further baselines from~\citep{lanctot2023population} that have not been trained against the evaluation population. Tabular Q-learning (QL) with recall length of 10 rounds and Contextual Regret Minimization (ContRM). For details we refer to previous work.

\subsection{Implementation Details}

We implement the CSRO oracle using the \emph{Gemini 2.5 Pro} model~\citep{comanici2025gemini25pushingfrontier}. 
All experiments are run for $K = 20$ iterations and $M = 10$ for \emph{LinearRefinement}. 
For the \emph{AlphaEvolve} oracle, we follow the method described in \citep{romera2024mathematical,novikov2025alphaevolve}. 
Results are averaged over 5 seeds if not specified otherwise.
The prompts and initial strategies are provided in Supplementary Materials~\ref{appendix:prompts}.



\section{Results}
\label{sec:results}

\begin{figure*}[t!]
    \centering
    \includegraphics[width=0.99\linewidth]{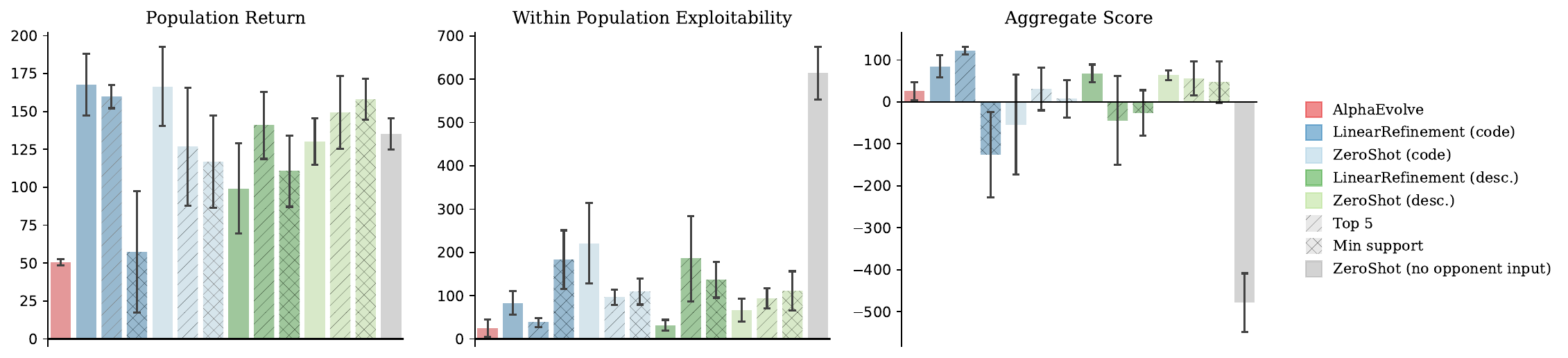}
    \caption{Performance of CSRO variants in Repeated Rock-Paper-Scissors}\label{fig:csro_rrps_performance}
\end{figure*}

\subsection{Repeated Rock-Paper-Scissors}

\begin{table}[!h]
\centering

    \begin{tabular}{
    l
    S[table-format=3.1(2.1),table-align-uncertainty,separate-uncertainty]
    S[table-format=3.1(2.1),table-align-uncertainty,separate-uncertainty]
    S[table-format=3.1(2.1),table-align-uncertainty,separate-uncertainty]
}
\toprule
\textbf{Method / Agent} & \textsc{PopReturn}~$\uparrow$ & \textsc{PopExpl}~$\downarrow$ & \textsc{AggScore}~$\uparrow$ \\
\midrule
CSRO & & & \\
- AlphaEvolve & 50.5 \pm 1.9 & 25.2 \pm 20.3 & 25.4 \pm 21.6 \\
- LinearRef. (code) & 159.8 \pm 7.7 & 37.7 \pm 10.6 & 122.1 \pm 9.8 \\
- LinearRef. (desc.) & 99.3 \pm 29.9 & 31.6 \pm 12.4 & 67.7 \pm 21.4 \\
- ZeroShot (desc.) & 130.2 \pm 15.4 & 66.7 \pm 25.9 & 63.5 \pm 11.4 \\
\midrule
LLM Agent (27B) & 193.2 & 67.2 & 126.0 \\
PSRO-IMPALA & -108.9 \pm 17.6 & 423.2 \pm 28.0 & -532.1 \pm 41.5 \\
\midrule
QL (R=10)$^{*}$ & -0.5 & 8.6 & -9.1 \\
ContRM$^{*}$ & 164.8 & 16.3 & 148.5 \\
LLM Agent (70B)$^{*}$ & 201.0 & 45.8 & \cellcolor{gray!25} 155.2 \\
\bottomrule
\end{tabular}
\vspace{.5em}\caption{Performance in Repeated Rock-Paper-Scissors across \emph{Population Return}, \emph{Within Population Exploitability} and \emph{Aggregate Score}. Results are averaged over 5 seeds. {\normalfont($^*$\,Denotes results reported in~\protect\cite{lanctot2023population}.)}}\label{tab:rrps_results}
\end{table}

\noindent Our primary results in RRPS are summarized in Table~\ref{tab:rrps_results}. We report \textsc{PopReturn} (higher is better), \textsc{PopExpl} (lower is better), and \textsc{AggScore} (higher is better) for the meta-equilibrium strategies of CSRO variants at the final iteration ($K=20$) as well as baseline performances. Our experiments show that among the proposed CSRO variants, the \emph{AlphaEvolve} oracle is the most effective at generating a low-exploitability policy with a population exploitability of \num{25.2 \pm 20.3}. 
This outcome is consistent with the overall objective of the PSRO framework. The oracle generates policies that minimize worst-case performance against the current strategy mixture. As weaker opponents are assigned less support in the meta-game equilibrium, the oracle's objective naturally prioritizes robustness against strong strategies over maximizing exploitation of potential weak opponents. A powerful evolutionary process like AlphaEvolve is particularly well-suited to drive this optimization.
\emph{LinearRefinement (code)} with \emph{Top 5} filtering achieves the highest aggregate score among our tested CSRO variants (\num{122.1 \pm 9.8}). This performance is competitive with the strongest baseline LLM agent, a 27B parameter Gemma 3 model (\num{126.0}).
We find that CSRO with intra-iteration policy refinement exhibits lower exploitability than the baseline Gemma 3 agents across all model sizes.
Furthermore, all CSRO variants substantially outperform the PSRO-IMPALA baseline across all three metrics.


For context within the broader literature, \citep{lanctot2023population} reported an impressive aggregate score of \num{155.2} for an LLM agent based on a Chinchilla model with 70B parameters, this is significantly higher than most baseline RL algorithms considered in their work. Among agents trained via self-play, ContRM, a contextual regret minimizer, achieved an aggregate score of \num{148.51}. Their findings also underscore the importance of a balanced evaluation; for instance, while the Q-learning agent with recall~10 achieved very low exploitability (\num{8.6}), its performance was undermined by a near-zero \textsc{PopReturn}, resulting in an \textsc{AggScore} lower than that of a bot playing uniformly at random (\texttt{randbot}).

A detailed overview of results for all CSRO variants is shown in Figure~\ref{fig:csro_rrps_performance} (and Supplementary Materials Table~\ref{tab:rrps_full_results}). 
The oracle's input format (code vs. description) interacts with the generation process. The benefit of using source code as input appears to be linked to iterative refinement. In the \emph{ZeroShot} setting, providing opponent policies as natural language descriptions leads to better results; \emph{ZeroShot (desc.)} achieves an aggregate score of \num{63.5 \pm 11.4}, compared to \num{-54.3 \pm 118.7} for \emph{ZeroShot (code)}. A potential explanation is that generating a best response from a large set of source code in a single pass is a more complex task for which a textual summary provides a more tractable input. As a critical ablation, the \emph{ZeroShot (no opponent input)} variant, which receives no information on opponent strategies, has a high \textsc{PopReturn} (\num{135.3 \pm 10.2}) but also an extremely high \textsc{PopExpl} (\num{614.2 \pm 60.8}), leading to a very poor aggregate  score (\num{-478.9 \pm 70.2 }). This result underscores that opponent conditioning is the most important component of the prompt for generating robust strategies.

The \emph{Top 5} filter generally yields better results than \emph{Min support}. A plausible explanation is that the meta-game equilibrium often concentrates its probability mass on the most recent, single-best counter-policy. In such cases, \emph{Min support} filtering would provide the LLM with only one opponent strategy as context. This can lead to overfitting, where the oracle generates a narrow policy that is highly effective against that one opponent but brittle against the wider, low-probability population. In contrast, the \emph{Top 5} filter ensures a more diverse set of opponents is always present in the prompt, encouraging the LLM to generate more general and robust policies.

\subsection{Repeated Leduc Hold’em Poker}

\begin{table}[!th]
\centering
    \begin{tabular}{
    l
    S[table-format=2.1(1.1)]
    S[table-format=2.1(1.1)]
    S[table-format=3.1(2.1)]
}
\toprule
\textbf{Method} & \textsc{PopReturn}~$\uparrow$ & \textsc{PopExpl}~$\downarrow$ & \textsc{AggScore}~$\uparrow$ \\
\midrule
CSRO & & & \\
- AlphaEvolve & 49.3 \pm 3.7 & 4.4 \pm 0.6 & \cellcolor{gray!25} 44.9 \pm 4.1 \\
- LinearRefinement & 43.8 \pm 2.5 & 9.8 \pm 3.0 & 34.0 \pm 4.3 \\
- ZeroShot & 40.4 \pm 1.6 & 19.6 \pm 2.1 & 20.7 \pm 3.0 \\
\midrule
PSRO-IMPALA & 13.3 \pm 6.9 & 58.4 \pm 3.3 & -45.0 \pm 10.1 \\
CFR+ & 39.8 \pm 0.3 & 0.0 \pm 0.0 & 39.8 \pm 0.3\\
\bottomrule
\end{tabular}

\vspace{.5em}\caption{Performance in repeated Leduc hold'em. Results are averaged over 3 seeds.}\label{tab:poker_exploitability}
\end{table}

\noindent In Table~\ref{tab:poker_exploitability}, we report CSRO performance in repeated Leduc poker as well as PSRO-IMPALA and CFR+ baselines, where metrics are computed with respect to the evaluation population {$\mathcal{P}$} consisting of CFR+, \emph{AlwaysCall} and \emph{AlwaysFold}.
Here we use description input and no filtering for all oracles, as this produced lowest exploitability in RRPS (see Supplementary Materials Table~\ref{tab:rrps_full_results}).
Results show that the \emph{AlphaEvolve} oracle achieves the highest \textsc{PopReturn} and \textsc{AggScore}, showing its efficacy of producing generalizable strategies.
CSRO-AlphaEvolve also achieves an exploitability of \num{4.4\pm 0.6}, competitive with CFR+, demonstrating the robustness of the meta-equilibrium found.
We found that the bots' worst-case performances (\textsc{PopExpl}) consistently arose from matches against the CFR+ opponent.
Both the average-case (\textsc{PopReturn}) and worst-case (\textsc{PopExpl}) performances of the CSRO variants directly correlates with the strength of the best-response oracle: \emph{AlphaEvolve}, followed by \emph{LinearRefinement} and \emph{ZeroShot}. 
This finding underscores the critical role of a powerful best-response solver in achieving game-theoretically near-optimal policies in repeated Leduc Poker within the CSRO framework.


We present a detailed breakdown of returns against each opponent in the eval population in Supplementary Materials Table~\ref{tab:leduc_raw_results}. Against \emph{AlwaysCall}, CSRO-AlphaEvolve achieves an average return of \num{110.3 \pm 9.7} by developing a powerful value betting strategy~(see Section~\ref{sec:qualitative_analysis}). This performance is significantly higher than that of the PSRO-IMPALA (\num{57.7 \pm 3.3}) and CFR+ (\num{62.1 \pm 0.8}) baselines.
This demonstrates a key advantage of the policies found by CSRO framework: CSRO does not just learn a low-exploitable strategy, but finds strategies that dynamically capitalize on the opponents' predictability over the course of the repeated game to achieve a super-Nash level of exploitation.
\emph{LinearRefinement} achieves a higher average return against \emph{AlwaysFold} (\num{57.3 \pm 8.8}) compared to \emph{AlphaEvolve} (\num{42.0 \pm 3.2}) and on par with CFR+ (\num{57.4\pm0.2}).
\emph{AlphaEvolve} minimizes exploitability against CFR+ while exploiting \emph{AlwaysCall} significantly more than CFR+. \emph{LinearRefinement} finds strategies that lose more to CFR+, still exploiting \emph{AlwaysCall} more than CFR+, and exploiting \emph{AlwaysFold} as much as CFR+.
In summary, we see evidence that the various CSRO oracles find different non-dominated strategies.

\subsection{Qualitative Analysis}
\label{sec:qualitative_analysis}

A key advantage of CSRO is the generation of interpretable policies, which allows for direct inspection of an agent's strategic logic. By analyzing the best programs from our experiments, we find they contain multiple strategic components, parameters and even heuristic value estimations.

\subsubsection{Repeated Rock-Paper-Scissors}
The best-performing strategy, generated by \emph{LinearRefinement (no filter) - Top 5}, is a sophisticated ensemble agent achieving an impressive \textsc{PopReturn} of \num{238.3} (see Supplementary Materials Table~\ref{tab:rrps_full_results}) that would have placed it in between 2nd and 3rd place in the competition. Its well-commented source code allows for a direct analysis of its strategy, which is built on several key principles.
First, the agent's strategy is a form of brute-force pattern matching, enabled by a comprehensive ensemble of 32  predictors. Rather than attempting to first deduce an opponent's high-level strategy (e.g., Markov player), the agent tests a wide array of hypotheses in parallel on every single turn. This large portfolio—which includes high-order Markov models, reactive models, and numerous heuristic detectors—is designed on the principle that one of its specialized experts will likely find a strong correlation in any given opponent's behavior, relying on breadth rather than deduction.
Second, it uses an aggressive, high-confidence adaptation mechanism. Expert votes are weighted by the fifth power of their score, allowing the agent to rapidly and decisively commit to a theory of the opponent's strategy based on recent success. This non-linear weighting enables fast adaptation in a competitive environment.
Finally, its most advanced component is an explicit second-order ``Theory of Mind'' model. This component infers the opponent's likely predictive model by observing which of its own experts is currently most successful. It then simulates the opponent's prediction of its own move and plays the corresponding counter-move. See Supplementary Materials Listing~\ref{lst:rrps_bot} for the full source code.

\subsubsection{Repeated Leduc hold'em poker}
The policy that achieves the best \textsc{PopReturn} (77.8) and best \textsc{AggScore} (69.1) demonstrates an effective synthesis of opponent modeling to drive a classic Expected Value (EV) calculation.
The EV is calculated based on two key estimations: the \emph{equity} (the probability of winning at showdown) and a \emph{statistical model} of the opponent's tendencies (e.g., the probability they will fold to a bet).

The agent combines these estimations to calculate the EV of a potential raise as a weighted average of two primary outcomes: the immediate profit from winning the pot if the opponent folds, versus the potential net gain or loss from playing for a larger pot at showdown if the opponent calls. This calculation explicitly weighs the risk and reward of bluffing versus betting for value.

The interpretability of this structure allows us to see precisely how it produces expert-level, adaptive strategies. For instance:

Against an \emph{AlwaysCall} opponent, the bot learns the opponent’s folding probability is near zero. Consequently, its EV calculation simplifies to a classic ``value betting'' strategy, where it ceases to bluff and only raises with very strong hands.
Against an \emph{AlwaysFold} opponent, the bot rapidly learns the folding probability is near 100\%. Its EV calculation for a raise then becomes equal to the current pot size, making its own hand irrelevant. As a result, the agent learns to bluff relentlessly to exploit the opponent's passivity.
This transparent adaptation, directly observable in the policy's code, demonstrates a level of strategic reasoning and interpretability that is fundamentally absent in opaque, black-box policies. See Supplementary Materials Listing~\ref{lst:leduc_poker_bot} for the full source code.

The generated policies share a key characteristic across both games and multiple seeds: they are not monolithic but rather compositions of distinct, interpretable strategic modules. Crucially, policies often demonstrate a capacity for sophisticated opponent modeling. This demonstrates a capacity for higher-order reasoning that is transparent and verifiable, a feature that distinguishes CSRO-generated policies from their black-box counterparts.

\vspace{-.5em}
\section{Related Work}
\label{sec:related_work}

Our work builds directly upon the Policy-Space Response Oracles (PSRO) framework~\citep{Lanctot2017PSRO}, a general-purpose algorithm for computing approximate Nash equilibria in large extensive-form games. Significant follow-up work has focused on improving the meta-solver, such as using alpha-rank~\citep{Omidshafiei2019AlphaRank} or developing more scalable training schemes like Pipeline PSRO~\citep{mcaleer2020pipelinepsro}. While these advancements have improved convergence and scalability, they have largely retained the use of deep reinforcement learning as the core of the best response oracle, thereby inheriting its limitations in terms of sample complexity and policy opacity. CSRO is orthogonal to these meta-solver improvements; we instead focus on fundamentally changing the nature of the oracle itself to prioritize interpretability.

Another line of work focuses on applying game-theoretic solvers to guide strategic interactions of LLMs in dialogue-based games~\citep{gemp2024steering}.
The authors formalize natural language negotiation and debate scenarios as extensive-form games. They integrate LLMs with solvers like CFR and PSRO to generate prompts containing high-level strategic instructions to the game-playing LLM, thereby steering its dialogue to be more rational and less exploitable. A key difference is that the best responses produced by their method are natural language prompts, whereas whereas CSRO outputs executable code. 
Previous work by Kempinski et al.~\citep{Kempinski2025GameOfThoughts} also explores using LLMs for iterative reasoning in game-theoretic settings. In their work, ``Game of Thoughts,'' they propose a family of algorithms inspired by cognitive hierarchy theory, such as level-k reasoning, to iteratively refine an LLM's strategic play. Their methods, including a variant named Policy Space Response Oracle Language Model (PSROLM), similarly place an LLM within an ``outer loop'' to improve its performance against prior strategies. While both approaches validate the use of LLMs for sophisticated strategic reasoning, our CSRO framework is distinct in two fundamental ways. First, our primary contribution is the synthesis of explicit, human-readable \emph{programmatic policies}. This focus on generating executable and commented code as the policy representation provides a level of interpretability and verifiability not present in methods that generate actions directly. Second, our algorithm incorporates a novel intra-iteration refinement loop, which tactically hardens a candidate policy against the current meta-game \emph{before} adding it to the policy pool, ensuring a higher quality of strategies.

Our work is most closely related to~\citep{bachrach2025combining} introducing LLM-PSRO. The authors focus on demonstrating the viability of LLMs as a code-generating, best-response oracle within a self-play loop. LLM-PSRO does not feature an inner feedback loop to iteratively refine the candidate policy. The work only includes inter-population evaluation and does not show performance against external populations or baselines.

\vspace{-.5em}
\section{Conclusion and Discussion}
\label{sec:conclusion}

We introduced Code-Space Response Oracles (CSRO), a novel framework for finding approximate equilibria in multi-agent games. We addressed the critical limitation of policy opacity in standard methods like PSRO by replacing the conventional black-box, deep RL oracle with a Large Language Model. Our approach reframes the best response computation as a program synthesis task, prompting an LLM to generate policies directly as human-readable, executable code.

The results show that CSRO achieves competitive performance in terms of convergence to a low-exploitability equilibrium, particularly in complex games like multi-hand Leduc hold'em poker. In RRPS, while the \textsc{AggScore} of CSRO is below that of the best-performing LLM-based baseline reported in previous work, it is important to highlight the significant difference in computational efficiency. The baseline LLM agents are invoked on every turn, i.e., requiring \num{1000} model calls for a single game. In contrast, \emph{LinearRefinement} generates a complete, reusable policy where the number of LLM calls only grows linearly with the number of iterations (e.g., $K=20$ in our experiments). This demonstrates a favorable trade-off between peak performance and the computational cost of policy generation, positioning CSRO as a practical and efficient method for strategy discovery.

We acknowledge that the LLM’s pre-training data likely contains knowledge of game strategies for a classic game like Rock-Paper-Scissors. 
However, the oracle must synthesize a novel best response to a specific, dynamically generated mixture of programmatic opponents provided in-context. The success of this process demonstrates a sophisticated capability for in-context strategic reasoning and code generation, not merely pattern retrieval.

Despite its promising results, our approach has several limitations. First, the performance of the CSRO oracle is inherently tied to the capabilities of the underlying LLM and the quality of the prompt. Poorly structured prompts or less capable models can lead to suboptimal or syntactically incorrect code, requiring error-handling and regeneration logic. Second, while our method reduces the high sample complexity associated with RL training, it introduces the computational cost of repeated LLM API calls, which can be significant. Finally, the scalability of CSRO to games with vast, high-dimensional observation spaces (e.g., Stratego or StarCraft) remains an open question. Representing complex state and opponent strategies within the context length of current LLMs remains a substantial engineering challenge.

\noindent\rule{\columnwidth}{0.1pt}
{\footnotesize\noindent This manuscript was created with assistance from Gemini 2.5 Pro. The tool was used for revision and editing to improve clarity. The authors have carefully reviewed and approved all changes.}

\newpage
\bibliography{references}

\clearpage
\onecolumn

\appendix
\newpage

\section{Supplementary materials}
\label{sec:appendix}
\subsection{Hyperparameters}

\subsubsection{PSRO-IMPALA}\label{sec:psro-impala}
For all of our experiments, we perform a hyperparameter sweep over learning rates, hidden layer sizes, unroll length, entropy cost, batch size, and max gradient norm (see~Table~\ref{tab:hyper}). Best performing hyper parameters are shown in Table~\ref{tab:impala-hyper-rrps} and Table~\ref{tab:impala-hyper-leduc}.

\begin{table}[!h]
\vspace{-.5em}
\centering
\begin{tabular}{lc}
\toprule
\textbf{Hyperparameter} & \textbf{Range} \\
\midrule
learning rates         & [$0.0001$, $0.001$]\\
hidden layer sizes & \{[$256$, $128$], [$512$, $256$]\}\\
unroll length  &  [$20$, $40$]\\
entropy cost         & [$0.001$, $0.1$]\\
batch size & [$16,32,64$]\\
max gradient norm & \{1, 10, 40\} \\
\bottomrule
\end{tabular}
\vspace{.5em}
\caption{IMPALA hyperparameter sweep.}\label{tab:hyper}\vspace{-.5em}
\end{table}

\begin{table}[!h]
\vspace{-.5em}
\centering
\begin{tabular}{lc}
\toprule
\textbf{Hyperparameter} & \textbf{Value} \\
\midrule
learning rates         & $0.0001$\\
hidden layer sizes & [$256$, $128$]\\
unroll length  &  $20$\\
entropy cost         & $0.001$\\
batch size & $16$\\
max gradient norm & 40 \\
observation representation & observation\_tensor() in OpenSpiel~\citep{lanctot2019openspiel} \\
\bottomrule
\end{tabular}
\vspace{.5em}
\caption{Hyperparameters in RRPS.}\label{tab:impala-hyper-rrps}\vspace{-.5em}
\end{table}

\begin{table}[!h]
\vspace{-.5em}
\centering
\begin{tabular}{lc}
\toprule
\textbf{Hyperparameter} & \textbf{Value} \\
\midrule
learning rates         & $0.001$\\
hidden layer sizes & [$512$, $256$]\\
unroll length  &  $40$\\
entropy cost         & $0.01$\\
batch size & $64$\\
max gradient norm & 40 \\
observation representation & infostate\_tensor() in OpenSpiel~\citep{lanctot2019openspiel} \\
\bottomrule
\end{tabular}
\vspace{.5em}
\caption{Hyperparameter for repeated Leduc hold'em poker.}\label{tab:impala-hyper-leduc}\vspace{-.5em}
\end{table}

\begin{table}[!ht]
\vspace*{\fill}
\centering
\begin{tabular}{lrrrr}
\toprule
& \multicolumn{4}{c}{\textbf{Gemma 3 (PT)}} \\
\textbf{Bot Names} & 270M & 1B & 4B & 27B \\
\midrule
\texttt{actr\_lag2\_decay} & -20.1 & -40.2 & -22.1 & -18.7 \\
\texttt{adddriftbot2} & 68.1 & 39.4 & 83.7 & 65.4 \\
\texttt{addshiftbot3} & 120.5 & 134.2 & 186.6 & 199.7 \\
\texttt{antiflatbot} & 995.3 & 993.1 & 991.1 & 991.2 \\
\texttt{antirotnbot} & 53.4 & 49.8 & 65.9 & 51.6 \\
\texttt{biopic} & -25.3 & -44.3 & -25.6 & -35.2 \\
\texttt{boom} & 27.6 & -1.4 & 4.4 & -21.2 \\
\texttt{copybot} & 965.6 & 962.4 & 968.2 & 969.8 \\
\texttt{debruijn81} & -34.3 & -45.4 & -14.1 & -7.4 \\
\texttt{driftbot} & 90.7 & 27.3 & 100.4 & 78.0 \\
\texttt{flatbot3} & 5.6 & 99.1 & 127.7 & 137.4 \\
\texttt{foxtrotbot} & 15.0 & 47.6 & 33.9 & 12.6 \\
\texttt{freqbot2} & 608.1 & 863.0 & 917.0 & 936.5 \\
\texttt{granite} & 116.8 & 79.2 & 84.9 & 99.7 \\
\texttt{greenberg} & -47.1 & -84.1 & -52.0 & -29.3 \\
\texttt{halbot} & 5.1 & -96.7 & -87.8 & -55.9 \\
\texttt{inocencio} & 522.1 & 405.5 & 293.2 & 568.8 \\
\texttt{iocainebot} & -80.2 & -131.2 & -68.1 & -35.6 \\
\texttt{marble} & 113.2 & 74.5 & 91.9 & 98.8 \\
\texttt{markov5} & -34.2 & -58.2 & -7.2 & -18.5 \\
\texttt{markovbails} & -39.2 & -39.6 & -6.2 & -21.2 \\
\texttt{mixed\_strategy} & 48.2 & 1.2 & 37.6 & 26.6 \\
\texttt{mod1bot} & -55.9 & -60.5 & -54.9 & -50.8 \\
\texttt{multibot} & 287.0 & 177.8 & 265.2 & 253.0 \\
\texttt{peterbot} & 650.6 & 770.1 & 865.6 & 879.6 \\
\texttt{phasenbott} & -85.2 & -141.3 & -64.8 & -52.8 \\
\texttt{pibot} & 6.5 & -36.0 & -21.5 & -28.5 \\
\texttt{piedra} & 33.6 & 22.5 & 40.6 & 40.6 \\
\texttt{predbot} & 46.6 & -38.1 & -27.5 & -67.2 \\
\texttt{r226bot} & 197.2 & 354.1 & 394.7 & 391.8 \\
\texttt{randbot} & 6.4 & 7.4 & 5.3 & -9.3 \\
\texttt{robertot} & -17.1 & -13.1 & -11.8 & -27.9 \\
\texttt{rockbot} & 998.4 & 998.4 & 995.6 & 995.0 \\
\texttt{rotatebot} & 978.0 & 980.9 & 988.8 & 997.5 \\
\texttt{russrocker4} & -51.8 & -71.5 & -31.2 & -38.9 \\
\texttt{shofar} & -33.6 & -44.2 & -24.8 & -34.8 \\
\texttt{sunCrazybot} & 248.0 & 313.7 & 371.6 & 397.9 \\
\texttt{sunNervebot} & -24.2 & -71.3 & -30.7 & -32.8 \\
\texttt{sweetrock} & 36.6 & 41.6 & 38.9 & 42.0 \\
\texttt{switchalot} & 67.0 & 106.9 & 111.9 & 125.2 \\
\texttt{switchbot} & 181.6 & 218.4 & 204.8 & 226.3 \\
\texttt{textbot} & 57.3 & 138.9 & 180.1 & 153.8 \\
\texttt{zq\_move} & 181.2 & 107.8 & 151.8 & 155.9 \\
\midrule
\textsc{PopReturn} & 167.0 & 162.7 & 187.2 & 193.2 \\
\textsc{PopExpl} & 85.2 & 141.3 & 87.8 & 67.2 \\
\textsc{AggScore} & 81.8 & 21.4 & 99.4 & 126.0 \\
\bottomrule
\end{tabular}
\vspace{.5em}
\caption{LLM agent performance against Repeated Rock-Paper-Scissors bot population. Returns averaged over 16 games.}\label{tab:gemma3_rrps}
\vspace*{\fill}
\end{table}

\begin{table*}[!t]
\centering
\resizebox{\columnwidth}{!}{
\begin{tabular}{lllS[table-format=4.1(2.1),table-align-uncertainty,separate-uncertainty]rS[table-format=4.1(2.1),table-align-uncertainty,separate-uncertainty]rS[table-format=4.1(3.1),table-align-uncertainty,separate-uncertainty]r}
\toprule
{\textbf{Method}} & {\textbf{Input}} & {\textbf{Filter}} & \multicolumn{2}{c}{{\textsc{PopReturn~$\uparrow$}}} & \multicolumn{2}{c}{{\textsc{PopExpl}~$\downarrow$}} & \multicolumn{2}{c}{{\textsc{AggScore}~$\uparrow$}} \\
{\textbf{}} & {\textbf{}} & {\textbf{}} & {\textbf{\quad~\quad~mean}} & {\textbf{max}} & {\textbf{\quad~\quad~mean}} & {\textbf{min}} & {\textbf{\quad~\quad~mean}} & {\textbf{max}} \\
\midrule
AlphaEvolve & Desc. & - & 50.5 \pm 1.9 & 54.2 & \cellcolor{gray!25} 25.2 \pm 20.3 & \cellcolor{gray!25} 3.3 & 25.4 \pm 21.6 & 47.6 \\
LinearRefinement & Desc. & - & 99.3 \pm 29.9 & 171.8 & 31.6 \pm 12.4 & 11.8 & 67.7 \pm 21.4 & 128.1 \\
LinearRefinement & Code & Top 5 & 159.8 \pm 7.7 & 189.1 & 37.7 \pm 10.6 & 8.8 & \cellcolor{gray!25} 122.1 \pm 9.8 & 141.3 \\
ZeroShot & Desc. & - & 130.2 \pm 15.4 & 177.8 & 66.7 \pm 25.9 & 18.9 & 63.5 \pm 11.4 & 86.9 \\
LinearRefinement & Code & - & \cellcolor{gray!25} 167.7 \pm 20.4 & \cellcolor{gray!25} 238.3 & 83.2 \pm 27.4 & 15.7 & 84.5 \pm 26.1 & 148.1 \\
ZeroShot & Desc. & Top 5 & 149.4 \pm 23.9 & 207.3 & 93.5 \pm 23.5 & 38.2 & 55.8 \pm 40.0 & \cellcolor{gray!25} 155.9 \\
ZeroShot & Code & Top 5 & 126.7 \pm 39.0 & 190.0 & 95.8 \pm 17.9 & 53.2 & 30.9 \pm 51.0 & 111.7 \\
ZeroShot & Code & Min support & 116.9 \pm 30.6 & 160.9 & 109.5 \pm 30.1 & 24.8 & 7.4 \pm 45.0 & 122.7 \\
ZeroShot & Desc. & Min support & 157.9 \pm 13.5 & 204.5 & 111.0 \pm 45.2 & 38.0 & 46.9 \pm 48.9 & 133.1 \\
LinearRefinement & Desc. & Min support & 110.6 \pm 23.5 & 185.9 & 137.0 \pm 41.6 & 28.9 & -26.4 \pm 54.2 & 84.4 \\
LinearRefinement & Code & Min support & 57.3 \pm 40.0 & 152.3 & 183.2 \pm 67.7 & 54.2 & -125.8 \pm 101.6 & 98.1 \\
LinearRefinement & Desc. & Top 5 & 140.9 \pm 22.2 & 187.6 & 185.2 \pm 98.5 & 67.1 & -44.3 \pm 105.7 & 110.5 \\
ZeroShot & Code & - & 166.6 \pm 26.1 & 208.3 & 220.9 \pm 92.7 & 95.8 & -54.3 \pm 118.7 & 112.6 \\
ZeroShot & - & - & 135.3 \pm 10.2 & 173.2 & 614.2 \pm 60.8 & 411.7 & -478.9 \pm 70.2 & -238.5 \\
\bottomrule
\end{tabular}
}
\vspace{.5em}
\caption{Performance of CSRO variants in Repeated Rock-Paper-Scissors ranked by \textsc{PopExpl}. Metrics are averaged over 3 seeds for AlphaEvolve and 5 seeds for all other variants.}\label{tab:rrps_full_results}
\end{table*}

\begin{table*}[!t]
\centering
\begin{tabular}{
    l
    S[table-format=4.1(2.1)]
    S[table-format=4.1(2.1)]
    S[table-format=4.1(2.1)]
}
\toprule
& \multicolumn{3}{c}{\textbf{Opponent}} \\
\textbf{Method} & \emph{CFR+} & \emph{AlwaysCall} & \emph{AlwaysFold} \\
\midrule
CSRO & & & \\
- AlphaEvolve  &  -4.4 \pm 0.6 & \cellcolor{gray!25} 110.3 \pm 9.7 &  42.0 \pm 3.2 \\
- LinearRefinement &  -9.8 \pm 3.0       &  83.8 \pm 5.1    & 57.3 \pm  8.8 \\
- ZeroShot  & -19.6 \pm 2.1       &  99.7 \pm 3.0    &  41.0 \pm  6.1 \\
\midrule
PSRO-IMPALA & -58.4 \pm 3.3 &   57.7 \pm 3.3 & 40.7 \pm 14.1 \\
CFR+ & \cellcolor{gray!25} 0.0 \pm 0.0 & 62.1 \pm 0.8 & \cellcolor{gray!25} 57.4\pm0.2 \\
\bottomrule
\end{tabular}

\vspace{.5em}
\caption{Performance of CSRO variants in repeated Leduc hold'em. Metrics are averaged over 3 seeds.}\label{tab:leduc_raw_results}
\end{table*}

\clearpage

\onecolumn

\begin{lstlisting}[
caption={A repeated Rock-Paper-Scissors bot that achieves best average performance against the opponent pool},
language=Python, 
label=lst:rrps_bot,
]
from collections import defaultdict
import random
from typing import Any, Callable, Dict, List, Optional, Tuple


class Agent:
  """An expert agent for Repeated Rock Paper Scissors, engineered to defeat
  a field of sophisticated, adaptive opponents.

  This agent's strategy is a significant enhancement of the 'ensemble of 
  experts' paradigm, built on four core principles:

  1.  **Superset Expert Ensemble**: The agent employs a vast portfolio of
      predictive experts designed to be a strict superset of those used by its
      competitors. This includes high-order Markov models (up to 8th order),
      complex reactive/joint models, and a wide range of periodic and
      heuristic predictors. By having more 'lenses' to view the game, it can
      detect patterns that other agents cannot.

  2.  **Aggressive, High-Confidence Adaptation**: The agent uses a highly
      aggressive weighted voting system, weighting each expert's vote by the
      *fifth power* of its score. This allows the agent to rapidly and
      decisively lock onto successful predictive models while effectively
      eliminating noise from the dozens of other, less-successful experts.

  3.  **Advanced Meta-Prediction ('Theory of Mind')**: The agent not only models
      the opponent but also models itself being modeled by the opponent. Its
      most critical expert, `meta_imitation`, deduces the opponent's likely
      predictive model by observing which of its own strategies is currently
      most successful. It then simulates the opponent's prediction of its own
      move and plays the perfect counter to the opponent's anticipated play.

  4.  **Strategic Randomization**: To combat the meta-predictors of its
      opponents, the agent introduces randomness at key decision points. When
      any of its predictive models or the final vote tally results in a tie,
      it chooses randomly among the best options. This makes its own behavior
      significantly harder to predict and exploit.

  Combined with finely tuned hyperparameters for a 1000-round game (a
  long-memory decay rate of 0.985), this agent is designed to out-learn, 
  out-adapt, and out-think its competition.
  """

  def __init__(self):
    """Initializes the agent's constants, models, and meta-strategy state."""
    # Game constants
    self.MOVES = ['ROCK', 'PAPER', 'SCISSORS']
    self.COUNTER_MOVES = {
        'ROCK': 'PAPER',
        'PAPER': 'SCISSORS',
        'SCISSORS': 'ROCK',
    }

    # Hyperparameters tuned for a 1000-round game against strong learners
    self.decay = 0.985  # Slower decay for longer memory and pattern stability
    self.vote_power = (
        5  # Use score^5 for extremely aggressive, high-confidence adaptation
    )

    # State is initialized in reset()
    self.reset()

  def reset(self) -> None:
    """Resets all agent state for the beginning of a new match."""
    self.my_history: List[str] = []
    self.opponent_history: List[str] = []

    # --- Underlying predictive models (Opponent's patterns) ---
    self.opp_markov_models = {
        i: defaultdict(lambda: defaultdict(int)) for i in range(1, 9)
    }  # n=1 to 8
    self.reactive_model = defaultdict(lambda: defaultdict(int))
    self.joint_hist_model = defaultdict(lambda: defaultdict(int))
    self.freq_model = defaultdict(int)

    # --- Underlying predictive models (My own patterns for meta-experts) ---
    self.my_markov_models = {
        i: defaultdict(lambda: defaultdict(int)) for i in range(1, 9)
    }  # n=1 to 8
    self.my_reactive_model = defaultdict(lambda: defaultdict(int))
    self.my_joint_hist_model = defaultdict(lambda: defaultdict(int))

    # --- Meta-learning state (The Experts) ---
    self.experts: Dict[str, Callable[[], Optional[str]]] = {
        # Opponent-history-based predictors (Markov up to 8th order)
        **{
            f'opp_markov{i}': self._create_opp_markov_predictor(i)
            for i in range(1, 9)
        },
        # My-history-based predictors
        'reactive': self._predict_reactive,
        'joint_hist': self._predict_joint_hist,
        # Simple statistical predictors
        'frequentist': self._predict_frequentist,
        # Heuristic predictors
        'copy_opponent': self._predict_copy_opponent,
        'copy_me': self._predict_copy_me,
        'rotator': self._predict_rotator,
        'counter_rotator': self._predict_counter_rotator,
        # Periodic predictors (up to 12 rounds)
        **{
            f'periodic_{i}': self._create_periodic_predictor(i)
            for i in range(2, 13)
        },
        # Standard Meta-predictors (Theory of Mind Level 1)
        **{
            f'meta_my_markov{i}': self._create_meta_my_markov_predictor(i)
            for i in range(1, 4)
        },
        'meta_my_reaction': self._predict_meta_my_reaction,
        'meta_my_joint': self._predict_meta_my_joint,
        # Advanced Meta-predictor (Theory of Mind Level 2)
        'meta_imitation': self._predict_meta_imitation,
    }
    self.expert_scores = {name: 1.0 for name in self.experts}
    self.last_predictions: Dict[str, Optional[str]] = {
        name: None for name in self.experts
    }

  # --- Update Methods ---

  def _update_scores(self, actual_opponent_move: str) -> None:
    """Updates expert scores based on their last prediction's accuracy."""
    for name, prediction in self.last_predictions.items():
      self.expert_scores[name] *= self.decay
      if prediction == actual_opponent_move:
        self.expert_scores[name] += 1

  def _update_models(self, my_action: str, opponent_action: str) -> None:
    """Updates all internal statistical models with the last round's data."""
    self.freq_model[opponent_action] += 1

    if self.my_history:
      self.reactive_model[self.my_history[-1]][opponent_action] += 1
      joint_key = (self.my_history[-1], self.opponent_history[-1])
      self.joint_hist_model[joint_key][opponent_action] += 1
      self.my_joint_hist_model[joint_key][my_action] += 1

    if self.opponent_history:
      self.my_reactive_model[self.opponent_history[-1]][my_action] += 1

    for n in self.opp_markov_models:
      if len(self.opponent_history) >= n:
        key = tuple(self.opponent_history[-n:])
        self.opp_markov_models[n][key][opponent_action] += 1
      if len(self.my_history) >= n:
        key = tuple(self.my_history[-n:])
        self.my_markov_models[n][key][my_action] += 1

  # --- Expert Prediction Methods ---

  def _get_best_prediction(self, model: Dict, key: Any) -> Optional[str]:
    """Helper to get the most likely move from a model given a key.

    Crucially, it breaks ties randomly to make our agent less predictable.
    """
    predictions = model.get(key)
    if not predictions:
      return None
    max_val = max(predictions.values())
    best_moves = [move for move, val in predictions.items() if val == max_val]
    return random.choice(best_moves)

  def _create_opp_markov_predictor(self, n: int) -> Callable[[], Optional[str]]:
    """Factory for creating n-th order Markov model predictors for the opponent."""

    def predictor() -> Optional[str]:
      if len(self.opponent_history) < n:
        return None
      key = tuple(self.opponent_history[-n:])
      return self._get_best_prediction(self.opp_markov_models[n], key)

    return predictor

  def _predict_reactive(self) -> Optional[str]:
    if not self.my_history:
      return None
    return self._get_best_prediction(self.reactive_model, self.my_history[-1])

  def _predict_joint_hist(self) -> Optional[str]:
    if not self.my_history:
      return None
    key = (self.my_history[-1], self.opponent_history[-1])
    return self._get_best_prediction(self.joint_hist_model, key)

  def _predict_frequentist(self) -> Optional[str]:
    return self._get_best_prediction({'freq': self.freq_model}, 'freq')

  def _predict_copy_opponent(self) -> Optional[str]:
    return self.opponent_history[-1] if self.opponent_history else None

  def _predict_copy_me(self) -> Optional[str]:
    return self.my_history[-1] if self.my_history else None

  def _predict_rotator(self) -> Optional[str]:
    if not self.opponent_history:
      return None
    return self.COUNTER_MOVES.get(self.opponent_history[-1])

  def _predict_counter_rotator(self) -> Optional[str]:
    if not self.opponent_history:
      return None
    return {v: k for k, v in self.COUNTER_MOVES.items()}.get(
        self.opponent_history[-1]
    )

  def _create_periodic_predictor(self, n: int) -> Callable[[], Optional[str]]:
    """Factory for creating predictors for various periodicities."""

    def predictor() -> Optional[str]:
      return (
          self.opponent_history[-n] if len(self.opponent_history) >= n else None
      )

    return predictor

  def _create_meta_my_markov_predictor(
      self, n: int
  ) -> Callable[[], Optional[str]]:
    """Assumes opponent predicts my n-th order markov pattern and counters it."""

    def predictor() -> Optional[str]:
      if len(self.my_history) < n:
        return None
      key = tuple(self.my_history[-n:])
      predicted_my_move = self._get_best_prediction(
          self.my_markov_models[n], key
      )
      return (
          self.COUNTER_MOVES.get(predicted_my_move)
          if predicted_my_move
          else None
      )

    return predictor

  def _predict_meta_my_reaction(self) -> Optional[str]:
    """Assumes opponent predicts my reaction pattern and counters it."""
    if not self.opponent_history:
      return None
    predicted_my_reaction = self._get_best_prediction(
        self.my_reactive_model, self.opponent_history[-1]
    )
    return (
        self.COUNTER_MOVES.get(predicted_my_reaction)
        if predicted_my_reaction
        else None
    )

  def _predict_meta_my_joint(self) -> Optional[str]:
    """Assumes opponent predicts my joint history pattern and counters it."""
    if not self.my_history:
      return None
    key = (self.my_history[-1], self.opponent_history[-1])
    predicted_my_move = self._get_best_prediction(self.my_joint_hist_model, key)
    return (
        self.COUNTER_MOVES.get(predicted_my_move) if predicted_my_move else None
    )

  def _predict_meta_imitation(self) -> Optional[str]:
    """Predicts by assuming the opponent is using a model similar to our own

    best-performing model to predict our moves. This is Theory of Mind Level 2.
    """
    # Define the set of our experts that an opponent is likely to be using.
    opponent_modeling_experts = {
        'opp_markov1': (self.my_markov_models[1], self.my_history, 1),
        'opp_markov2': (self.my_markov_models[2], self.my_history, 2),
        'opp_markov3': (self.my_markov_models[3], self.my_history, 3),
        'reactive': (self.my_reactive_model, self.opponent_history, 1),
        'joint_hist': (
            self.my_joint_hist_model,
            (self.my_history, self.opponent_history),
            2,
        ),
    }
    # Find which of these opponent-modeling experts is performing best for us.
    best_expert_name = max(
        opponent_modeling_experts,
        key=lambda name: self.expert_scores.get(name, 0),
    )
    model, history_data, hist_len = opponent_modeling_experts[best_expert_name]

    key = None
    # Construct the key for the corresponding model of *our* behavior.
    if best_expert_name == 'joint_hist':
      if len(history_data[0]) >= 1 and len(history_data[1]) >= 1:
        key = (history_data[0][-1], history_data[1][-1])
    elif isinstance(history_data, list) and len(history_data) >= hist_len:
      key = tuple(history_data[-hist_len:])

    if key is None:
      return None

    # Predict our own move using the model the opponent is likely using.
    predicted_my_move = self._get_best_prediction(model, key)
    # Assume the opponent will play the counter to our predicted move.
    return (
        self.COUNTER_MOVES.get(predicted_my_move) if predicted_my_move else None
    )

  # --- Main Agent Logic ---

  def act(self, observation: dict[str, Any]) -> str:
    """Determines the agent's next move by learning from the past and using the

    weighted ensemble to predict the opponent's action.
    """
    # First round: reset state and play randomly.
    if observation.get('my_action') is None:
      self.reset()
      return random.choice(self.MOVES)

    # --- 1. Learning Phase (from previous turn's outcome) ---
    my_prev_action = observation['my_action']
    opp_prev_action = observation['opponent_action']

    self._update_scores(opp_prev_action)
    self._update_models(my_prev_action, opp_prev_action)
    self.my_history.append(my_prev_action)
    self.opponent_history.append(opp_prev_action)

    # --- 2. Prediction Phase (for the current turn) ---
    for name, func in self.experts.items():
      self.last_predictions[name] = func()

    # --- 3. Action Phase (for the current turn) ---
    vote_tally = defaultdict(float)
    for name, prediction in self.last_predictions.items():
      if prediction is not None:
        vote_tally[prediction] += self.expert_scores[name] ** self.vote_power

    if vote_tally:
      # Determine the most likely opponent move using our tie-breaking helper.
      predicted_opponent_move = self._get_best_prediction(
          {'vote': vote_tally}, 'vote'
      )
      return self.COUNTER_MOVES[predicted_opponent_move]

    # Ultimate fallback: If no expert could predict, play randomly.
    return random.choice(self.MOVES)
\end{lstlisting}

\newpage
\begin{lstlisting}[language=Python, caption={A repeated Leduc Poker bot that achieves best \textsc{PopReturn} against the opponent pool}, label=lst:leduc_poker_bot]
from typing import Any, Dict


class RepeatedLeducPokerBot:
  """A bot using a non-stationary strategy.

  It collapses its strategy into a single, continuous 'bravery' parameter. This
  parameter adapts based on the opponent's observed passivity and aggression,
  creating a reactive, non-stationary strategy designed to be difficult for
  learning opponents to model.
  """

  def __init__(self):
    """Initializes the bot."""
    self.player_id: int = -1
    self.opponent_id: int = -1
    self.card_ranking = {'J': 1, 'Q': 2, 'K': 3}
    self.game_count = 0

    self.alpha = 1.0  # Laplace smoothing parameter.
    # Model: info_set_str -> {'FOLD': count, 'CALL': count, 'RAISE': count}
    self.opponent_model = {}

  def _get_info_set_key(self, round_name, public_card, round_history) -> str:
    """Creates a unique key for an information set."""
    public_card_str = public_card if public_card else ''

    action_str = ''.join(a['action'][0] for a in round_history)
    return f'{round_name}:{public_card_str}:{action_str}'

  def receive_outcome(self, obs: Dict[str, Any]):
    """Receives the game outcome and updates the opponent action model."""
    self.game_count += 1
    action_history = obs.get('action_history', {})

    for round_name in ['PREFLOP', 'POSTFLOP']:
      round_history = action_history.get(round_name, [])
      # We need to build the history incrementally to get the info set *before* each action
      temp_round_history = []
      for action_event in round_history:
        if action_event.get('player_id') == self.opponent_id:
          public_card = (
              obs['public_state'].get('public_card')
              if round_name == 'POSTFLOP'
              else None
          )
          info_set_key = self._get_info_set_key(
              round_name, public_card, temp_round_history
          )

          if info_set_key not in self.opponent_model:
            self.opponent_model[info_set_key] = {
                'FOLD': self.alpha,
                'CALL': self.alpha,
                'RAISE': self.alpha,
            }

          action = action_event.get('action')
          if action in self.opponent_model[info_set_key]:
            self.opponent_model[info_set_key][action] += 1

        temp_round_history.append(action_event)

  def restart(self, player_id: int):
    """Starts a new game, assigning player position and adapting 'bravery'.

    Args:
        player_id: The player ID (0 or 1) for the new game.
    """
    self.player_id = player_id
    self.opponent_id = 1 - player_id

    # This method is called at the start of a new game.
    # We can reset any per-game state here if needed.
    pass

  def act(self, obs: Dict[str, Any]) -> str:
    """Outputs an action based on Expected Value (EV) calculations."""
    legal_actions = obs['player_view']['legal_actions']
    if not legal_actions:
      return 'CALL'

    # 1. Calculate Equity
    my_hand = obs['player_view']['hand']
    public_card = obs['public_state']['public_card']
    equity = self._calculate_equity(my_hand, public_card, obs)  # Pass obs here

    # 2. Calculate EV for each legal action
    evs = {}
    for action in legal_actions:
      evs[action] = self._calculate_action_ev(action, obs, equity)

    # 3. Choose best action
    best_action = max(evs, key=evs.get)
    return best_action

  def _calculate_equity(
      self, my_hand: str, public_card: str or None, obs: Dict[str, Any]
  ) -> float:
    """Calculates the equity of our hand against a distribution of opponent hands,

    weighted by opponent's observed actions in the current hand.
    Equity is the probability of winning at showdown.
    """
    deck = ['J', 'J', 'Q', 'Q', 'K', 'K']
    deck.remove(my_hand)
    if public_card:
      if public_card in deck:
        if public_card in deck:  # Check membership before removing
          deck.remove(public_card)

    possible_opponent_hands = deck
    if not possible_opponent_hands:
      return 0.5  # No possible hands for opponent, return neutral equity

    opponent_hand_weights = {
        hand: 1.0 for hand in possible_opponent_hands
    }  # Start with uniform weights

    # --- Add Weighting based on History ---
    action_history = obs.get('action_history', {})
    preflop_history = action_history.get('PREFLOP', [])
    postflop_history = action_history.get('POSTFLOP', [])

    opponent_preflop_raises = 0
    opponent_preflop_calls = 0  # Calls implying >0 bet (heuristic)
    opponent_postflop_raises = 0
    opponent_postflop_calls = 0  # Calls implying >0 bet (heuristic)

    # Simplified action parsing (doesn't know exact bet amounts)
    temp_round_hist = []
    for action_event in preflop_history:
      if action_event['player_id'] == self.opponent_id:
        action = action_event['action']
        # Check if this action faces a bet based on prior actions
        facing_bet = (
            any(
                a['action'] == 'RAISE'
                for a in temp_round_hist
                if a['player_id'] == self.player_id
            )
            or self.player_id == 1
        )  # P2 faces BB

        if action == 'RAISE':
          opponent_preflop_raises += 1
        elif (
            action == 'CALL' and facing_bet
        ):  # Count calls that match a bet/raise
          opponent_preflop_calls += 1
      temp_round_hist.append(action_event)

    temp_round_hist = []
    # Base postflop bet state on whether preflop ended with aggression
    postflop_facing_bet = False  # Initial state for P1 postflop
    if preflop_history:
      last_preflop_action = preflop_history[-1]
      # If last preflop action was a raise (either ours or theirs that we called)
      if last_preflop_action['action'] == 'RAISE':
        postflop_facing_bet = (
            True  # Simplified: assumes aggression carries over
        )

    for action_event in postflop_history:
      if action_event['player_id'] == self.opponent_id:
        action = action_event['action']
        # Check if opponent faced a bet in this round before their action
        currently_facing_bet = any(
            a['action'] == 'RAISE'
            for a in temp_round_hist
            if a['player_id'] == self.player_id
        )

        if action == 'RAISE':
          opponent_postflop_raises += 1
          postflop_facing_bet = True  # Their raise makes us face a bet
        elif (
            action == 'CALL' and currently_facing_bet
        ):  # Count calls that match a bet/raise
          opponent_postflop_calls += 1
          postflop_facing_bet = False  # Call equalizes bets
        elif (
            action == 'CALL' and not currently_facing_bet
        ):  # Zero-cost call (check)
          postflop_facing_bet = False  # Check equalizes bets
      else:  # Our action
        if action_event['action'] == 'RAISE':
          postflop_facing_bet = True  # We raised, opponent faces bet
        elif action_event['action'] == 'CALL':  # Our check or call
          # If we called a raise, bets are equal. If we checked, bets are equal.
          postflop_facing_bet = False
      temp_round_hist.append(action_event)

    # Apply weights (Heuristics - Requires Tuning)
    if opponent_preflop_raises > 0:
      for hand in list(opponent_hand_weights.keys()):
        if hand == 'J':
          opponent_hand_weights[hand] *= 0.1  # Less likely J
        if hand == 'Q':
          opponent_hand_weights[hand] *= 0.7
        if hand == 'K':
          opponent_hand_weights[hand] *= 2.0  # More likely K
    elif opponent_preflop_calls > 0:  # Called preflop but didn't raise
      for hand in list(opponent_hand_weights.keys()):
        if hand == 'J':
          opponent_hand_weights[hand] *= 1.2
        if hand == 'K':
          opponent_hand_weights[hand] *= 0.5  # Less likely K

    if public_card:
      if opponent_postflop_raises > 0:
        for hand in list(opponent_hand_weights.keys()):
          opp_rank_tuple = self._get_hand_rank(hand, public_card)
          is_pair = (
              opp_rank_tuple[0] >= 3 + self.card_ranking['J']
          )  # Check if it's any pair
          if not is_pair:
            opponent_hand_weights[
                hand
            ] *= 0.1  # Unlikely to raise postflop without a pair
          else:
            opponent_hand_weights[
                hand
            ] *= 2.0  # More likely to have a pair if raised

    # Normalize weights to form a probability distribution
    total_weight = sum(opponent_hand_weights.values())
    if (
        total_weight <= 1e-9
    ):  # Avoid division by zero / handle case where all weights -> 0
      # Reset to uniform if weights collapse
      opponent_hand_weights = {hand: 1.0 for hand in possible_opponent_hands}
      total_weight = sum(opponent_hand_weights.values())
      if total_weight <= 1e-9:
        return 0.5  # Still no weight, return neutral equity

    opponent_hand_probs = {
        hand: weight / total_weight
        for hand, weight in opponent_hand_weights.items()
    }
    # Filter out hands with zero probability for efficiency
    opponent_hand_probs = {
        h: p for h, p in opponent_hand_probs.items() if p > 1e-9
    }

    if not opponent_hand_probs:
      return 0.5  # Check again after filtering

    win, tie = 0.0, 0.0  # Use floats for weighted sum
    total_prob_considered = 0.0

    if public_card:  # Postflop Equity Calculation
      my_rank = self._get_hand_rank(my_hand, public_card)
      for opp_hand, prob in opponent_hand_probs.items():
        opp_rank = self._get_hand_rank(opp_hand, public_card)
        if my_rank > opp_rank:
          win += prob
        elif my_rank == opp_rank:
          tie += prob
        total_prob_considered += prob

      if total_prob_considered == 0:
        return 0.5
      # Weighted Equity = (Sum P(OppHand)*Win(OppHand) + 0.5 * Sum P(OppHand)*Tie(OppHand)) / Sum P(OppHand)
      # Since Sum P(OppHand) = total_prob_considered, this simplifies:
      return (win + 0.5 * tie) / total_prob_considered

    else:  # Preflop Equity Calculation (Average over public cards)
      total_weighted_outcomes = 0.0  # Sum of probabilities of paths considered

      # Need the deck *after* my card is removed, before considering opponent hand
      base_deck_for_opp = ['J', 'J', 'Q', 'Q', 'K', 'K']
      base_deck_for_opp.remove(my_hand)

      for opp_hand, opp_prob in opponent_hand_probs.items():
        # Create the deck for board cards given my_hand and opp_hand
        deck_for_board = list(base_deck_for_opp)
        if opp_hand not in deck_for_board:
          # This can happen if opponent_hand_weights included hands now impossible
          # due to filtering or deck state issues. Skip this opponent hand.
          continue
        deck_for_board.remove(opp_hand)

        if not deck_for_board:  # No possible board cards (e.g., K vs K preflop)
          # Rank based on private card only
          my_rank_priv = self.card_ranking[my_hand]
          opp_rank_priv = self.card_ranking[opp_hand]
          outcome_weight = (
              opp_prob  # Weight by probability of this opponent hand
          )
          if my_rank_priv > opp_rank_priv:
            win += outcome_weight
          elif my_rank_priv == opp_rank_priv:
            tie += outcome_weight
          total_weighted_outcomes += outcome_weight
          continue  # Skip board card loop for this opponent hand

        # Loop through possible board cards
        num_board_cards = len(deck_for_board)
        board_card_prob = (
            1.0 / num_board_cards
        )  # Uniform prob for each possible board card

        for board_card in deck_for_board:
          outcome_weight = (
              opp_prob * board_card_prob
          )  # Combined probability of this path

          my_rank_board = self._get_hand_rank(my_hand, board_card)
          opp_rank_board = self._get_hand_rank(opp_hand, board_card)

          if my_rank_board > opp_rank_board:
            win += outcome_weight
          # Loss contributes 0 to 'win' or 'tie'
          elif my_rank_board == opp_rank_board:
            tie += outcome_weight
          total_weighted_outcomes += outcome_weight

      if total_weighted_outcomes == 0:
        return 0.5  # Default if something went wrong
      # Equity = (Weighted Wins + 0.5 * Weighted Ties) / Total Weight Considered
      # Note: total_weighted_outcomes should ideally sum close to 1.0 if all paths are covered.
      # Normalizing by total_weighted_outcomes accounts for any paths skipped (e.g., impossible opp hands)
      return (win + 0.5 * tie) / total_weighted_outcomes

  def _get_hand_rank(
      self, private_card: str, public_card: str or None
  ) -> tuple[int, int]:
    """Assigns a numeric rank tuple to a hand for comparison."""
    private_rank = self.card_ranking[private_card]
    if not public_card:
      # Pre-showdown, just rank by private card
      return (private_rank, 0)

    public_rank = self.card_ranking[public_card]
    # Postflop
    if private_rank == public_rank:  # Pair
      # Major rank for pairs (4,5,6) is higher than for high cards (1,2,3)
      return (3 + private_rank, 0)
    else:  # High card
      # Major rank is the highest card, minor rank is the kicker
      return (max(private_rank, public_rank), min(private_rank, public_rank))

  def _get_opponent_action_probs(self, obs: Dict[str, Any]) -> Dict[str, float]:
    """Gets the opponent's likely response probabilities to our RAISE from our model."""
    round_name = obs['public_state']['round']
    public_card = obs['public_state']['public_card']

    current_history = obs['action_history'][round_name]
    # Simulate our raise to create the info set the opponent would face
    simulated_history = current_history + [
        {'player_id': self.player_id, 'action': 'RAISE'}
    ]
    info_set_key = self._get_info_set_key(
        round_name, public_card, simulated_history
    )

    if info_set_key in self.opponent_model:
      counts = self.opponent_model[info_set_key]
      total = sum(counts.values())
      if total > 0:
        return {action: count / total for action, count in counts.items()}

    # Default assumption if we have no data
    return {'FOLD': 0.33, 'CALL': 0.34, 'RAISE': 0.33}

  def _calculate_action_ev(
      self, action: str, obs: Dict[str, Any], equity: float
  ) -> float:
    """Calculates the Expected Value of a single action."""
    pot_size = obs['public_state']['pot_size']

    committed = [(100 - c) for c in obs['public_state']['chips']]
    my_committed = committed[self.player_id]
    opp_committed = committed[self.opponent_id]
    amount_to_call = opp_committed - my_committed

    round_name = obs['public_state']['round']
    raise_amount = 2 if round_name == 'PREFLOP' else 4

    if action == 'FOLD':
      return 0

    if action == 'CALL':
      pot_after_call = pot_size + amount_to_call
      return (equity * pot_after_call) - amount_to_call

    if action == 'RAISE':
      # EV(Raise) = P(Opponent Folds) * (Win Pot) + P(Opponent Calls) * (EV at Showdown)
      # We simplify by ignoring opponent re-raises.

      opp_probs = self._get_opponent_action_probs(obs)
      prob_opp_folds = opp_probs['FOLD']

      # What we win if opponent folds
      win_on_fold = pot_size

      # What happens if opponent calls our raise
      amount_to_raise = amount_to_call + raise_amount
      # Pot includes current pot, my raise, and opponent's call of the raise amount
      pot_after_opp_calls = pot_size + amount_to_raise + raise_amount

      value_if_called = (equity * pot_after_opp_calls) - amount_to_raise

      return (prob_opp_folds * win_on_fold) + (
          (1 - prob_opp_folds) * value_if_called
      )

    return -999  # Should not be reached
\end{lstlisting}

\subsection{Prompts and Initial Strategies}\label{appendix:prompts}

\subsubsection{Repeated Rock-Paper-Scissors}
\begin{lstlisting}[caption={Prompt for Repeated Rock-Paper-Scissors}, label=lst:rrps_prompt, language=]
Act as an expert Python programmer and game theory enthusiast.

Your task is to implement an agent class called `Agent` that represents a
player's strategy in a game of Repeated Rock Paper Scissors.

## Game Rules:

* The game is played for 1000 rounds.
* Your goal is to maximize your total number of wins over the 1000 rounds.
* The valid moves are strings: 'ROCK', 'PAPER', 'SCISSORS'.
* ROCK beats SCISSORS, SCISSORS beats PAPER, and PAPER beats ROCK.

## Class Signature:

Your implementation must align with the following class structure.

```python
class Agent:

  def act(self, observation: dict[str, Any]) -> str:
    ...
```

## Input details:

The input `observation` is a dictionary with two keys representing the moves
played by the player and the opponent in the previous round. Defaults to `None`
for the first round.

```python
observation = {
  'my_action': 'ROCK',  # None for the first round.
  'opponent_action': 'PAPER',  # None for the first round.
}
```
\end{lstlisting}

\subsubsection{Repeated Leduc Poker}

\begin{lstlisting}[caption={Prompt for Repeated Leduc Poker}, label=lst:leduc_poker_prompt, language=]
Act as an expert in computer games, opponent modeling, algorithm design and multiagent learning. Your task is to iteratively improve the provided bot in repeated leduc poker. The bot will play leduc poker with an opponent for multiple games, where in each game their player positions are randomly permutated. The primary goal is to increase the scores on the provided evaluation metrics, where larger values are better.

Always adhere to best practices in Python coding.

# Rule of Leduc Poker
Leduc Poker is a simplified two-player poker game, ideal for AI research, that uses a small deck to focus on core poker concepts like betting strategy and imperfect information.

Here is a detailed breakdown of the rules to clarify legal moves. Note that in this implementation, the "Check" action is not available; players must use "Call" instead. A call may be zero-cost if there is no outstanding bet to match.

**1. Setup & Preliminaries**
*   **Players:** 2.
*   **Deck:** 6 cards (two Jacks, two Queens, two Kings).
*   **Blinds:** Before cards are dealt, mandatory bets are posted:
    *   Player 1 (P1) posts a **Small Blind** of 1 unit.
    *   Player 2 (P2) posts a **Big Blind** of 2 units.
*   **The Deal:** Each player receives one private card, face down.

**2. Core Betting Rules**
*   **Raise Sizing:** The amount to raise is fixed.
    *   **Round 1:** The raise amount is **2 units**.
    *   **Round 2:** The raise amount is **4 units**.
*   **Total Betting Cap:** The total betting cap for each round is a maximum of **two raises**.
*   **Acting First:** Player 1 (the small blind) acts first in both betting rounds (pre-flop and post-flop).

**3. Round 1: Pre-Flop Betting**
This round occurs before the public card is revealed.

*   **P1's First Action:** P1 must act on P2's 2-unit Big Blind.
    *   **Fold:** Forfeit the 1-unit blind. P2 wins the pot.
    *   **Call:** Match the 2 units by putting in 1 more unit.
    *   **Raise:** Make a 2-unit raise, for a total of 4 units (P1 puts in 3 units). The total betting cap has been reached.
*   **P2's Action:**
    *   If P1 **called**, P2 can **Call** (a zero-cost action, as bets are equal) to end the round, or **Raise** (by putting in 2 more units to make it 4 total).
    *   If P1 **raised**, P2 can only **Call** (by putting in 2 more units) or **Fold**. The betting cap has been reached.
*   **P1's Second Action (if necessary):** If P1 called and P2 then raised, the action returns to P1. P1 can only **Call** (by putting in 2 more units) or **Fold**.

**4. The Flop: Public Card**
After Round 1 betting concludes, one public card is dealt face-up. This card is shared by both players.

**5. Round 2: Post-Flop Betting**
This round occurs after the flop. There are no blinds.

*   **P1's First Action:**
    *   **Call:** Make a zero-cost call to pass the turn (as there is no outstanding bet).
    *   **Raise:** Make a 4-unit raise.
*   **P2's Action:**
    *   If P1 **called** (at zero-cost), P2 can also **Call** (at zero-cost, ending the round) or **Raise** 4 units.
    *   If P1 **raised**, P2 can **Call** (matching the 4 units), **Raise** (by putting in another 4 units, for a total bet of 8), or **Fold**. The total betting cap has been reached.
*   **Subsequent Actions:**
    *   If P2 **raised** (after P1's initial zero-cost call), the action returns to P1, who can **Call** (the 4 unit bet), **Raise** (to 8 total), or **Fold**. The total betting cap has been reached.
    *   If a player **raises**, the other player can only **Call** or **Fold**, as the betting cap has been reached.

**6. Showdown & Hand Ranking**
If neither player folds, a showdown occurs after Round 2 betting.

*   **Hand:** A player's hand is their private card combined with the public card.
*   **Hand Ranks (best to worst):**
    1.  **Pair:** Two cards of the same rank (e.g., J-J). Higher pairs beat lower pairs.
    2.  **High Card:** If no one has a pair, the player with the highest card wins (K > Q > J).
*   **Ties:** If both players have the same hand rank (e.g., both have a King-high), the pot is split.

**7. Winning**
A player wins the pot either by being the only one left after the other folds, or by having the best hand at showdown.


# Program Skeleton
You should design the bot according to the following APIs:

```python
class RepeatedLeducPokerBot:

  def receive_outcome(self, obs: dict[str, Any]):
    """Receive game outcome of previous game."""

  def restart(self, player_id: int):
    """Start a new round of leduc poker with being assigned the player position player_id."""
  
  def act(self, obs: dict[str, Any]) -> str:
    """Output an action given an observation.

      Args:
        obs: a JSON observation dictionary.
      Output:
        action, an action in {{'FOLD', 'CALL', 'RAISE'}}
    """
```

# Example observation format
The act method is called each time the agent need to make a decision during a game, given an observation in the format of JSON. Here are some examples of an observation:
{{'player_view': {{'player_id': 0, 'current_player': True, 'hand': 'K', 'legal_actions': ['CALL', 'RAISE']}}, 'public_state': {{'round': 'PREFLOP', 'chips': [99, 99], 'pot_size': 2, 'public_card': None}}, 'action_history': {{'PREFLOP': [], 'POSTFLOP': []}}, 'game_result': None}}

{{'player_view': {{'player_id': 1, 'current_player': True, 'hand': 'K', 'legal_actions': ['CALL', 'RAISE']}}, 'public_state': {{'round': 'POSTFLOP', 'chips': [99, 99], 'pot_size': 2, 'public_card': 'Q'}}, 'action_history': {{'PREFLOP': [{{'player_id': 0, 'action': 'CALL'}}, {{'player_id': 1, 'action': 'CALL'}}], 'POSTFLOP': [{{'player_id': 0, 'action': 'CALL'}}]}}, 'game_result': None}}

The receive_outcome method is called at the end of each game. The agent receives an observation of the final state of the game to improve game play for future games with the same opponent. The observation here contains final payoff for each player, and players' hands if the last action is not FOLD. Some examples:

{{'player_view': {{'player_id': 0, 'current_player': False, 'hand': 'K', 'legal_actions': []}}, 'public_state': {{'round': 'PREFLOP', 'chips': [101, 99], 'pot_size': 0, 'public_card': None}}, 'action_history': {{'PREFLOP': [{{'player_id': 0, 'action': 'RAISE'}}, {{'player_id': 1, 'action': 'FOLD'}}], 'POSTFLOP': []}}, 'game_result': {{'outcome': 'FOLD', 'returns': [1, -1], 'showdown_hands': None}}}}


{{'player_view':{{'player_id': 0, 'current_player': False, 'hand': 'J', 'legal_actions': []}}, 'public_state': {{'round': 'POSTFLOP', 'chips': [95, 105], 'pot_size': 0, 'public_card': 'Q'}}, 'action_history': {{'PREFLOP': [{{'player_id': 0, 'action': 'RAISE'}}, {{'player_id': 1, 'action': 'RAISE'}}, {{'player_id': 0, 'action': 'CALL'}}], 'POSTFLOP': [{{'player_id': 0, 'action': 'CALL'}}, {{'player_id': 1, 'action': 'CALL'}}]}}, 'game_result': {{'outcome': 'SHOWDOWN', 'returns': [-5, 5], 'showdown_hands': [{{'player_id': 0, 'hand': 'J'}}, {{'player_id': 1, 'hand': 'K'}}]}}}}

The restart method is called at the beginning of each game, where the agent is informed the player position it will act as in the next game.

legal_actions is a subset of {{'FOLD', 'CALL', 'RAISE'}}


# Current program
Here is the current program we are trying to improve (you will need to propose a modification to it below):

{code}

# Opponents
Here are the summary of opponent codes you are trying to beat:
{instances}

Try to reason about these opponents, and come up with a strategy that can exploit them in general.

# *SEARCH/REPLACE block* Rules:

Every *SEARCH/REPLACE block* must use this format:
1. The opening fence: ```python
2. The start of search block: <<<<<<< SEARCH
3. A contiguous chunk of up to 4 lines to search for in the existing source code
4. The dividing line: =======
5. The lines to replace into the source code
6. The end of the replace block: >>>>>>> REPLACE
7. The closing fence: ```

Every *SEARCH* section must *EXACTLY MATCH* the existing file content, character for character, including all comments, docstrings, etc.

*SEARCH/REPLACE* blocks will replace *all* matching occurrences.
Include enough lines to make the SEARCH blocks uniquely match the lines to change.

Keep *SEARCH/REPLACE* blocks concise.
Break large *SEARCH/REPLACE* blocks into a series of smaller blocks that each change a small portion of the file.
Include just the changing lines, and a few surrounding lines if needed for uniqueness.
Do not include long runs of unchanging lines in *SEARCH/REPLACE* blocks.

To move code within a file, use 2 *SEARCH/REPLACE* blocks: 1 to delete it from its current location, 1 to insert it in the new location.

Make sure that the changes you propose are consistent with each other. For example, if you refer to a new config variable somewhere, you should also propose a change to add that variable.

Example:
```python
<<<<<<< SEARCH
    return total_loss
=======
    # Add sparsity-promoting regularization to the loss.
    total_loss += self.hypers.l1_reg_weight * l1_reg

    return total_loss
{replace}
```
and
```python
<<<<<<< SEARCH
  return hyper.zipit([
=======
  return hyper.zipit([
      hyper.uniform('l1_reg_weight', hyper.interval(0.0, 0.01)),
{replace}
```

{lazy_prompt}
ONLY EVER RETURN CODE IN A *SEARCH/REPLACE BLOCK*!

# Task
{task_instruction} {focus_sentence} {trigger_chain_of_thought}
Describe each change with a *SEARCH/REPLACE block*.
\end{lstlisting}

The initial strategy is a heuristic strategy by zero-shot prompting "give a strong heuristic repeated Leduc Poker strategy":

\subsubsection{Repeated Leduc Poker}

\begin{lstlisting}[caption={Prompt for Repeated Leduc Poker}, label=lst:leduc_poker_initial]
import random
from typing import Any, Dict


class RepeatedLeducPokerBot:
  """
  A simple heuristic bot for repeated Leduc Poker.

  The bot's strategy is based on a set of rules that consider the strength
  of its private card, the public card (if revealed), and the current
  betting round. It does not learn from the opponent's behavior.
  """

  def __init__(self):
    """Initializes the bot."""
    self.player_id: int = -1
    self.card_ranking = {'J': 1, 'Q': 2, 'K': 3}

  def receive_outcome(self, obs: Dict[str, Any]):
    """
    Receives the game outcome of the previous game.

    For this simple heuristic bot, this method does nothing as it
    does not adapt its strategy based on past games.
    """
    pass

  def restart(self, player_id: int):
    """
    Starts a new game of Leduc Poker, assigning the player position.

    Args:
        player_id: The player ID (0 or 1) for the new game.
    """
    self.player_id = player_id

  def act(self, obs: Dict[str, Any]) -> str:
    """
    Outputs an action given an observation.

    Args:
        obs: A dictionary containing the game state from the player's perspective.
    
    Returns:
        A string representing the chosen action: 'FOLD', 'CALL', or 'RAISE'.
    """
    player_view = obs.get('player_view', {})
    public_state = obs.get('public_state', {})
    legal_actions = player_view.get('legal_actions', [])

    if not legal_actions:
      # If there are no legal actions, we can't do anything.
      # This can happen at the end of a hand.
      return 'CALL'

    hand = player_view.get('hand')
    round_name = public_state.get('round')

    if round_name == 'PREFLOP':
      return self._preflop_strategy(hand, legal_actions)
    elif round_name == 'POSTFLOP':
      public_card = public_state.get('public_card')
      return self._postflop_strategy(hand, public_card, legal_actions)

    # Default action if something unexpected happens
    return 'CALL' if 'CALL' in legal_actions else random.choice(legal_actions)

  def _preflop_strategy(self, hand: str, legal_actions: list[str]) -> str:
    """Determines the action for the pre-flop round."""
    hand_strength = self.card_ranking.get(hand, 0)

    # Strong hand (King) - play aggressively
    if hand_strength == 3:  # King
      if 'RAISE' in legal_actions:
        return 'RAISE'
      return 'CALL'

    # Medium hand (Queen) - play cautiously
    elif hand_strength == 2:  # Queen
      # If we can call, we do. Avoid folding if possible.
      if 'CALL' in legal_actions:
        return 'CALL'
      # This case should ideally not be hit if CALL is always an option
      # when FOLD is, but as a fallback.
      return 'FOLD'

    # Weak hand (Jack) - play defensively
    else: # Jack
      # We want to see the flop cheaply.
      if 'CALL' in legal_actions:
        return 'CALL'
      return 'FOLD'

  def _postflop_strategy(self, hand: str, public_card: str, legal_actions: list[str]) -> str:
    """Determines the action for the post-flop round."""
    hand_strength = self.card_ranking.get(hand, 0)
    public_card_strength = self.card_ranking.get(public_card, 0)
    
    # Check for a pair (strongest hand)
    if hand == public_card:
      if 'RAISE' in legal_actions:
        return 'RAISE'
      return 'CALL'

    # High card strategy
    # If our private card is better than the public card, we have a decent high card hand
    if hand_strength > public_card_strength:
      if 'RAISE' in legal_actions:
        return 'RAISE'
      return 'CALL'

    # If our private card is weaker than the public card, our hand is weak.
    else:
      # Check-call if possible (zero-cost call)
      # A zero-cost call is implied if we just need to call 0 chips.
      # We will simplify and just check/call.
      if 'CALL' in legal_actions:
        return 'CALL'
      return 'FOLD'
\end{lstlisting}

\end{document}